\documentclass[10pt,conference]{IEEEtran}

\usepackage{enumitem}
\usepackage{subfigure,cite,epsfig,dblfloatfix}
\usepackage[cmex10]{amsmath}
\usepackage{amssymb,mathrsfs,graphicx}
\usepackage{psfrag}

\let\oldbrace\{
\def\{{\oldbrace\kern0.5pt}

\newcommand{\Ac}{\mathcal{A}}

\newcommand{\markov}{\mathrel\multimap\joinrel\mathrel-\mspace{-9mu}\joinrel\mathrel-}

\newtheorem{theorem}{Theorem}

\newtheorem{remark}{Remark}

\newtheorem{corollary}{Corollary}

\begin{document}

\title{A New Converse Bound for Coded Caching}
\author{
\IEEEauthorblockN{Chien-Yi Wang, Sung Hoon Lim, and Michael Gastpar} \\
\IEEEauthorblockA{School of Computer and Communication Sciences\\EPFL\\ Lausanne,
Switzerland\\
\{chien-yi.wang, sung.lim, michael.gastpar\}@epfl.ch}
}

\maketitle

\renewcommand{\thefootnote}{\fnsymbol{footnote}}
\renewcommand{\thefootnote}{}

\begin{abstract}
An information-theoretic lower bound is developed for the caching system studied by Maddah-Ali and Niesen. By comparing the proposed lower bound with the decentralized coded caching scheme of Maddah-Ali and Niesen, the optimal memory--rate tradeoff is characterized to within a multiplicative gap of $4.7$ for the worst case, improving the previous analytical gap of $12$. Furthermore, for the case when users' requests follow the uniform distribution, the multiplicative gap is tightened to $4.7$, improving the previous analytical gap of $72$. As an independent result of interest, for the single-user average case in which the user requests multiple files, it is proved that caching the most requested files is optimal.
\end{abstract}

\section{Introduction}

Recently, Maddah-Ali and Niesen considered the following problem setup of {\em caching}: A file server has access to a database of $N$ files. There are $K$ users, each equipped with an individual cache of the same size. Each user wishes to retrieve one of the $N$ files. During the placement phase, some information is stored in the users' caches. During the delivery phase, the server sends updates through a shared link so that each user can recover the desired file from the cache content and the received update message. The goal is to design the cache contents and the update such that given a fixed cache memory, the update rate is minimized. For a given cache size, the maximum update rate required over all possible requests (the worst case) was studied in \cite{Maddah-Ali:14}. For the case when the users' requests follow some probability distribution, the averaged update rate (the average case) was studied in \cite{Niesen:14}.

For the worst case, Maddah-Ali and Niesen gave an analytical characterization of the optimal memory--rate tradeoff to within a multiplicative gap of $12$ \cite[Theorem $3$]{Maddah-Ali:14}. For the average case in which the requests are distributed independently and uniformly (the uniform case), they also gave an analytical characterization of the optimal memory--rate tradeoff to within a multiplicative gap of $72$ \cite[Claim $1$]{Niesen:14}. Later on, Zhang, Lin, and Wang \cite{Zhang:15} improved their arguments and gave a universal multiplicative-plus-additive gap (87$R^\star$+2) for the general average case. Other improved converse bounds can be found in \cite{Chen:15, Ghasemi:15, Ji15}.

In this work, we propose an information-theoretic lower bound for the average case. By comparing with the achievable memory--rate tradeoff in \cite[Theorem 1]{Maddah-Ali:15}, we tighten the multiplicative gap from $12$ to $4.7$ for the worst case (Theorem \ref{thm:worst}). Moreover, for the uniform case we tighten the multiplicative gap from $72$ to $4.7$ (Theorem \ref{thm:uniform}). 
As an essential lemma for the developed lower bound, and as an independent result of interest, we prove that caching the most requested files is an optimal caching strategy for the single-user average case where the user may request {\em multiple} files, i.e., any subset of the $N$ files (Theorem \ref{thm:subset_ind}).

{\em Notation:} We use calligraphic symbols (e.g., $\mathcal{X}$) to denote sets. Denote by $|\cdot|$ the cardinality of a set. Random variables and their realizations are represented by uppercase letters (e.g., $X$) and lowercase letters (e.g., $x$), respectively. The probability distribution of a random variable $X$ is denoted by $p_X$. We say that $X\markov Y \markov Z$ form a Markov chain if $p_{X,Y,Z} = p_Yp_{X|Y}p_{Z|Y}$. 

We denote $x^+ := \max\{x,0\}$ for all $x\in\mathbb{R}$ and $[a] := \{1,2,\cdots a\}$ for all $a\in\mathbb{N}$. Also, we denote $[0:a] = \{0\}\cup[a]$. Given any sequence or tuple $(x_1,x_2,\cdots,x_k)$ and any subset $\mathcal{J}\subset [k]$, we use two short-hand notations $x^\mathcal{J}$ and $x_\mathcal{J}$ for the subsequence $(x_i:i\in\mathcal{J})$. For the case $\mathcal{J}=[k]$, $x^{[k]}$ is simply denoted by $x^{k}$ or by $\mathbf{x}$. 

\begin{figure}[t!]
\begin{center}
\includegraphics[scale=0.75]{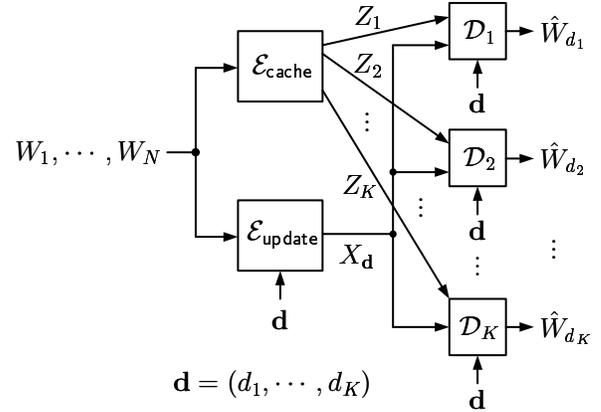}
\end{center}
\vspace{-0.15in}
\caption{The $K$-user caching problem with a database of $N$ files.}
\label{fig:system}
\vspace{-0.1in}
\end{figure}

\section{Problem statement} \label{sec:prob_state}
Denote by $N$ the number of files and by $K$ the number of users. Let $W_1,W_2\cdots,W_N$ be $N$ random variables independently and uniformly drawn from $[2^{F}]$, where $F$ is a positive integer. Each $W_n$ represents a file of size $F$ bits. On the other hand, we denote by $d_k$ the request of user $k\in[K]$. For notational convenience, we denote $W=(W_1,\cdots,W_N)$, $\mathbf{d}=(d_1,\cdots,d_K)$, and $\mathbf{R}=(R_\mathbf{d} : \mathbf{d}\in[N]^K)$. The $K$-user caching system is depicted in Figure \ref{fig:system}. 

An $(M,\mathbf{R})$ caching scheme consists of 
\begin{itemize}[leftmargin=*]
\item one cache encoder, which assigns $K$ indices $z_k(w)\in[2^{\lfloor FM \rfloor}]$, where $k\in[K]$, to each tuple $w\in[2^F]^N$;
\item one update encoder, which assigns an index $x_\mathbf{d}(w)\in[2^{\lfloor FR_\mathbf{d} \rfloor}]$ to each tuple $(w,\mathbf{d})\in[2^F]^N\times [N]^K$;
\item $K$ decoders, where decoder $k\in[K]$ maps the received messages and the requests, i.e., $(z_k,x_\mathbf{d},\mathbf{d})$, into an estimate $\hat{w}_{d_k}\in[2^F]$.
\end{itemize}
During the placement phase, the cache encoder maps the files $W_1,\cdots,W_N$ into the cache contents $Z_1,\cdots,Z_K$ and places
$Z_k$ in the cache of user $k\in[K]$. Then, during the delivery phase, the update encoder broadcasts the update message $X_\mathbf{d}$ to all users through the shared link. Finally, each user $k\in[K]$ recovers the desired file $\hat{W}_{d_k}$ from the received messages $(Z_k,X_\mathbf{d})$ and the requests $\mathbf{d}$. The probability of error is defined as 
\begin{IEEEeqnarray*}{rCl} 
{\sf P}_{\sf e} &:=& \max_{\mathbf{d}\in[N]^K}\max_{k\in[K]} \mathbb{P}(\hat{W}_{d_k}\neq W_{d_k}).
\end{IEEEeqnarray*}
We say that a rate tuple $(M,\mathbf{R})$ is {\em achievable} if for every $\epsilon>0$, there exists an $(M,\mathbf{R})$ caching scheme with large enough file size $F$ such that ${\sf P}_{\sf e}<\epsilon$. The optimal rate region $\mathcal{R}^\star$ is the closure of the set of achievable rate tuples. Given a fixed memory size $M\ge 0$, we restrict attention to the following projections of $\mathbf{R}$. The first projection is the maximum of update rates (thus the worst case): 
\begin{IEEEeqnarray}{rCl} 
R_{\sf worst}^\star(M) &:=& \min_{\mathbf{R}: (M,\mathbf{R})\in\mathcal{R}^\star} \max_{\mathbf{d}\in[N]^K} R_\mathbf{d}. \label{eq:worst-rate}
\end{IEEEeqnarray}
Denote by $p_\mathbf{D}$ a probability distribution of users' requests. The second projection is the weighted sum of update rates with weights $p_\mathbf{D}$ (thus the average case): 
\begin{IEEEeqnarray}{rCl} 
R_{\sf avg}^\star(M) &:=& \min_{\mathbf{R}: (M,\mathbf{R})\in\mathcal{R}^\star} \sum_{\mathbf{d}\in[N]^K} p_\mathbf{D}(\mathbf{d})R_\mathbf{d}. \label{eq:ave-rate}
\end{IEEEeqnarray}
In this work we assume that the requests $D_1,\ldots, D_K$ are i.i.d. drawn from the distribution $p_D$, i.e., $p_{\mathbf{D}}(\mathbf{d})=\prod_{k=1}^K p_{D}(d_k)$. When we specialize~\eqref{eq:ave-rate} to the uniform case, i.e., $p_D(d) = \frac{1}{N}$ for all $d\in[N]$, we denote the corresponding optimal memory--rate tradeoff by $R_{\sf uniform}^\star(M)$. Note that the uniform case models the scenario where the requests of different users are independent and the files are equally popular. Clearly, it holds that $R^\star_{\sf uniform}(M) \le R^\star_{\sf worst}(M)$ for all $M\ge 0$.

%
%

\section{Main Results} \label{sec:main}
Our first contribution is the following closed-form converse bound for the average case.
Without loss of generality, we assume that $p_D(1)\ge p_D(2)\ge \cdots \ge p_D(N)$. 
\begin{theorem} \label{thm:avg}
Consider the caching problem for the average case with request distribution $p_D$. For all $M\ge 0$, 
\begin{IEEEeqnarray*}{rCl}
R_{\sf avg}^\star(M) &\ge& \max_{k\in[K]} \sum_{n=1}^N (s_n(k)-s_{n+1}(k))(n-kM)^+,  
\end{IEEEeqnarray*}
where $s_{N+1}(k)=0$ and $s_n(k)=1-(1-p_D(n))^k$, $n\in[N]$.
\end{theorem}
\smallskip
The proof of this theorem is deferred to Section \ref{sec:lower}.
By setting $p_D(n) = \frac{1}{N}$, $n\in[N]$, in Theorem~\ref{thm:avg}, we have the converse bound for the uniform case which also serves as a converse bound for the worst case. 

\begin{corollary} \label{cor:uni}
Consider the caching problem for the uniform case. For all $M\ge 0$, 
\begin{IEEEeqnarray}{rCl} \label{eq:lower_uni}
R_{\sf uniform}^\star(M) &\ge& \max_{k\in[K]} \left(1-(1-1/N)^k\right)(N-kM)^+.
\end{IEEEeqnarray}
\end{corollary}
\smallskip

We will compare the proposed lower bound in Corollary \ref{cor:uni} with the decentralized coded caching scheme of Maddah-Ali and Niesen~\cite{Maddah-Ali:15}. The achievable memory--rate tradeoff is stated in the following theorem. 
\begin{theorem}[Maddah-Ali--Niesen~\cite{Maddah-Ali:15}] \label{thm:decentralized}
Consider the caching problem for the worst case. It holds that $R^\star_{\sf worst}(0) \le \min\{K,N\}$, and for all $M\in(0,N]$, 
\begin{IEEEeqnarray*}{ll} 
& R^\star_{\sf worst}(M) \\
\label{eq:MNdec}
&\le \left(N-M\right)\cdot \min\left\{\frac{1}{M}\left(1-\left(1-\frac{M}{N}\right)^K\right),1\right\} \IEEEyesnumber  \\
&=: R_{\sf MN}(K,N,M). 
\end{IEEEeqnarray*}
\end{theorem}
The achievable memory--rate tradeoff described in \eqref{eq:MNdec} is not convex. Thus, by time sharing among the achievable points, the achievable memory--rate tradeoff can be improved. We denote by $\breve{R}_{\sf MN}(K,N,M)$ the corresponding convexified bound. 

Our second contribution is in showing that given a fixed memory size $M\ge0$, the decentralized coded caching with time sharing achieves an update rate to within a constant multiplicative factor from the optimum memory--rate tradeoff $R^\star_{\sf worst}(M)$. The main result is the following theorem. 
\begin{theorem} \label{thm:worst}
For all $N\ge 1$, $K\ge 1$, and $M\in[0,N)$, 
\begin{IEEEeqnarray*}{rCl}
\frac{\breve{R}_{\sf MN}(K,N,M)}{R^\star_{\sf worst}(M)} &<& 4.7.
\end{IEEEeqnarray*}
\end{theorem}
Since our information-theoretic converse bound is developed for the general average case, we can also establish the following stronger claim.  
\begin{theorem} \label{thm:uniform}
For all $N\ge 1$, $K\ge 1$, and $M\in[0,N)$, 
\begin{IEEEeqnarray*}{rCl}
\frac{\breve{R}_{\sf MN}(K,N,M)}{R^\star_{\sf uniform}(M)} &<& 4.7.
\end{IEEEeqnarray*}
\end{theorem}

\begin{remark}
In \cite{Maddah-Ali:14}, Maddah-Ali and Niesen proposed the following lower bound for the worst case: 
\begin{IEEEeqnarray}{rCl} \label{eq:MNlower}
R_{\sf worst}^\star(M) &\ge& \max_{k\in[\min\{K,N\}]} \left(k - \frac{k}{\lfloor N/k\rfloor}M\right)^+,
\end{IEEEeqnarray}
for all $M\ge 0$.
Numerical evaluation reveals that not surprisingly, the lower bound \eqref{eq:MNlower} (which applies only to the worst case) is often tighter than the proposed lower bound \eqref{eq:lower_uni} (which applies both to the worst and to the uniform case). Nevertheless, there exist cases where \eqref{eq:lower_uni} is tighter than \eqref{eq:MNlower}, e.g., when $N=K=5$ and $M=1$.
\end{remark}

%
%

\section{Converse Bound for the Average Case} \label{sec:lower}
In this section, we present the average case converse bound. When attributing a distribution $p_\mathbf{D}$ on the requests, we further assume that $W$ and $\mathbf{D}$ are independent. 

For $n\in[N]$, we denote by $(B_{n1},B_{n2},\cdots,B_{nF})$ the binary representation of $W_n$. Since $W_n$ is uniformly distributed over $[2^F]$, $B_{n[F]}$ are i.i.d. Bernoulli($1/2$) random variables. For notational convenience, we denote $B=(B_1,\cdots,B_N)$, where the entries are i.i.d. Bernoulli($1/2$) random variables. Then, we have the following converse bound. 
\begin{theorem}\label{thm:dms-outer} 
Consider the caching problem for the average case with request distribution $p_ \mathbf{D}$. For all $M\ge 0$, 
\begin{IEEEeqnarray*}{rCl}
R_{\sf avg}^\star(M) &\ge& \min \max_{\mathcal{A}\subseteq [K]} H(B_{D_\mathcal{A}}|V_\mathcal{A}, \mathbf{D}),  
\end{IEEEeqnarray*}
where $B_{D_\mathcal{A}} = (B_{D_k}:k\in\mathcal{A})$ and the minimum is over all conditional pmfs $p_{V_{[K]}|B}$ such that $V_{[K]}\markov B \markov  \mathbf{D}$ form a Markov chain and 
\begin{IEEEeqnarray*}{rCl}
I(B;V_{\mathcal{A}}) &\le& |\mathcal{A}| M, 
\end{IEEEeqnarray*}
for all subsets $\mathcal{A}\subseteq[K]$.
\end{theorem}
\begin{IEEEproof}
Consider any subset $\mathcal{A}\subseteq [K]$. Recall that $B^F$ is the binary representation of $W$. Denote $V_{ki} = (Z_k,B^{i-1})$, $k\in[K]$, $i\in[F]$. Since $W$ and $\mathbf{D}$ are independent by assumption, the Markov chain $V_{[K]i}\markov B_i \markov \mathbf{D}$ holds for all $i\in[F]$. 
Then, since $H(Z_k) \le FM$ for all $k\in[K]$, we have 
\begin{IEEEeqnarray*}{rCl}
|\mathcal{A}| FM &\ge& \sum_{k\in \mathcal{A}} H(Z_k) \\
&\ge& H(Z_{\mathcal{A}}) \\
&=& I(W;Z_{\mathcal{A}}) \\
&=& I(B^F;Z_{\mathcal{A}}) \\
&=& \sum_{i=1}^F I(B_i;Z_{\mathcal{A}}|B^{i-1}) \\
&=& \sum_{i=1}^F I(B_i;V_{\mathcal{A}i}).
\end{IEEEeqnarray*}
Next, we have 
\begin{IEEEeqnarray*}{ll}
& F\sum_{\mathbf{d}\in[N]^K} p_{ \mathbf{D}}(\mathbf{d})R_\mathbf{d} \\
&\ge \sum_{\mathbf{d}\in[N]^K} p_{ \mathbf{D}}(\mathbf{d}) H(X_\mathbf{d}| \mathbf{D}=\mathbf{d}) \\
&= H(X_ \mathbf{D}| \mathbf{D}) \\
&\ge H(X_ \mathbf{D}|Z_{\mathcal{A}}, \mathbf{D}) \\
&= H(W_{D_{\mathcal{A}}},X_ \mathbf{D}|Z_{\mathcal{A}}, \mathbf{D}) - H(W_{D_{\mathcal{A}}}|X_ \mathbf{D},Z_{\mathcal{A}}, \mathbf{D}) \\
&\overset{(a)}{\ge} H(W_{D_{\mathcal{A}}}|Z_{\mathcal{A}}, \mathbf{D}) - F\epsilon_F \\
&= H(B^F_{D_{\mathcal{A}}}|Z_{\mathcal{A}}, \mathbf{D}) - F\epsilon_F \\
&= \sum_{i=1}^F H(B_{D_{\mathcal{A}}i}|B_{D_{\mathcal{A}}}^{i-1},Z_{\mathcal{A}}, \mathbf{D}) - F\epsilon_F \\
&\ge \sum_{i=1}^F H(B_{D_{\mathcal{A}}i}|V_{\mathcal{A}i}, \mathbf{D}) - F\epsilon_F 
\end{IEEEeqnarray*}
where $(a)$ follows from the data processing inequality and Fano's inequality, and $\epsilon_F$ tends to zero as $F\to \infty$. The rest of the proof follows from the standard time sharing argument and then letting $F\to\infty$. 
\end{IEEEproof}

Now let us restrict attention to the case of i.i.d. requests, i.e., $p_{ \mathbf{D}}(\mathbf{d}) = \prod_{k=1}^K p_D(d_k)$ for some distribution $p_D$ defined on $[N]$. 
Then, by symmetry, the bound in Theorem~\ref{thm:dms-outer}  only depends on the cardinality of $\mathcal{A}$. 
Furthermore, to facilitate the analysis, we relax the lower bound by swapping the positions of minimum and maximum. Then, we have the following corollary. 
\begin{corollary} \label{col:avg}
Consider the caching problem for the average case with $p_{ \mathbf{D}}(\mathbf{d}) = \prod_{k=1}^K p_D(d_k)$ for some distribution $p_D$. For all $M\ge 0$, 
\begin{IEEEeqnarray}{rCl} \label{eq:uni}
R_{\sf uniform}^\star(M) &\ge& \max_{k\in[K]} \min H(B_{D_{[k]}}|V_k,\mathbf{D}),  
\end{IEEEeqnarray}
where given a fixed $k\in[K]$, the minimum is over all conditional pmfs $p_{V_k|B}$ such that $V_k\markov B \markov \mathbf{D} $ form a Markov chain and $I(B;V_k) \le kM$.
\end{corollary}

Next, we give a closed-form expression of \eqref{eq:uni} by relating it to a single-user caching problem.
The caching network that we consider for this task is a generalization of our caching problem with $K=1$. In particular, we formulate a single-user caching system in which the user may request {\em multiple} files, namely, any subset of the $N$ files, with request distribution $p_Y$, where $Y$ is an element of the power set $\mathcal{P}([N])$. We refer to this caching network as the single-user multiple request caching network.

The relation between the multi-user single request caching network (our primary problem of interest) and the single-user multiple request caching network is as follows. In the multi-user single request setup, users $1,\cdots,K$ wish to recover files $W_{D_1},\cdots,W_{D_K}$, respectively. Following a cut-set based argument in which we assume that users in some subset $\Ac\subseteq[K]$ cooperate, the cache memories are combined resulting in a single cache of size $|\Ac|M$. Moreover, the (cooperative) decoder wishes to recover multiple files $(W_{D_k}: k\in\Ac)$. Thus, the optimal memory--rate tradeoff for the single-user multiple request network with memory size $|\Ac|M$ and request $Y=(D_k: k\in\Ac)$ serves as a lower bound on the multi-user single request network with memory size $M$ and requests $(D_1,\cdots,D_K)$. 

\subsection{Single-user multiple request caching} \label{sec:single_ind}
For each $n\in[N]$, we denote 
\begin{IEEEeqnarray}{rCl} \label{eq:sn}
s_n &:=& \mathbb{P}(Y\ni n) = \sum_{\substack{ y\in\mathcal{P}([N])  \\  \text{ s.t. } n\in y}} p_Y(y).
\end{IEEEeqnarray} 
Without loss of generality, we assume that $s_1\ge s_2 \ge \cdots \ge s_N$. 
We establish the following theorem for the single-user multiple request caching network.

\begin{theorem}\label{thm:subset_ind}
Consider the single-user multiple request caching problem with request distribution $p_Y$. The optimal memory--rate tradeoff for the average case is 
\begin{IEEEeqnarray*}{rCl}
R^\star_{\sf avg}(M) &=& \sum_{n=1}^N (s_n-s_{n+1}) \left(n-M\right)^+, 
\end{IEEEeqnarray*}
where $s_n$ is defined in \eqref{eq:sn} for all $n\in[N]$ and $s_{N+1}=0$.  
\end{theorem}
The proof of Theorem~\ref{thm:subset_ind} is deferred to Appendix. 
\begin{remark}
Theorem~\ref{thm:subset_ind} indicates that an optimal caching strategy for the single-user multiple request caching network is to cache the most popular files, where the popularity is measured by how often they are requested.
\end{remark}

\subsection{Proof of Theorem~\ref{thm:avg}}
Next, we observe that given any fixed $k\in[K]$, it holds that for all $n\in[N]$,  
\begin{IEEEeqnarray*}{rCl}
s_n &=& \mathbb{P}(n\in\{D_1,\cdots,D_k\}) \\
&=& 1 - \mathbb{P}(n\notin\{D_1,\cdots,D_k\}) \\
&=& 1 - \prod_{j=1}^k \mathbb{P}(D_j\neq n) \\
&=& 1 - (1-P_D(n))^k.
\end{IEEEeqnarray*}
Finally, by applying Theorem~\ref{thm:subset_ind} to Corollary~\ref{col:avg}, we have the closed-form converse bound in~Theorem~\ref{thm:avg}.

%
%

\section{The Gap Analysis: Proof of Theorems \ref{thm:worst} and \ref{thm:uniform}} \label{sec:analysis}
If $N=1$, it can be easily checked that $\breve{R}_{\sf MN}(K,1,M)=1-M=R^\star_{\sf uniform}(M)$. For $N\in\{2,3,4\}$, we have 
\begin{IEEEeqnarray*} {rCl}
\frac{\breve{R}_{\sf MN}(K,N,M)}{R^\star_{\sf uniform}(M)} &\le& \frac{N-M}{\max_{k\in[K]}(1-(1-1/N)^k)(N-kM)} \\
&\overset{k=1}{\le}& \frac{N-M}{(1-(1-1/N))(N-M)} \\
&=& N < 4.7.
\end{IEEEeqnarray*}
For the rest of analysis, we assume that $N\ge 5$. 
To facilitate the gap analysis, we consider the following relaxed upper bound of Theorem \ref{thm:decentralized}:
\begin{IEEEeqnarray*}{rCl} 
R_{\sf worst}(M) &\le& (N-M) \cdot  \min\left\{\frac{1}{M},1\right\} \\
&=:& R_{\sf upper}(K,N,M), 
\end{IEEEeqnarray*}
for all $M \in(0,N]$, and we define $R_{\sf upper}(K,N,0) := \min\{K,N\}$. We remark that $R_{\sf upper}(K,N,M)$ is quite suboptimal as an upper bound and is not continuous at $M=0$ when $K<N$. However, the corresponding convexified bound $\breve{R}_{\sf upper}(K,N,M)$ is sufficient for our analysis. On the other hand, we consider the following relaxed lower bound
\begin{IEEEeqnarray*}{ll} 
& R_{\sf uniform}^\star(M) \\
&\ge \max_{k\in[\min\{K,\lceil N/4\rceil\}]}(1-(1-1/N)^k)(N-kM)^+ \\
&=: R_{\sf lower}(K,N,M). 
\end{IEEEeqnarray*}
Since $R_{\sf lower}(K,N,M) \le R_{\sf uniform}^\star(M) \le R_{\sf worst}^\star(M) \le \breve{R}_{\sf upper}(K,N,M)$, it suffices to show 
\begin{align*}
\frac{\breve{R}_{\sf upper}(K,N,M)}{R_{\sf lower}(K,N,M)} < 4.7, \quad M\in[0,N).
\end{align*} 
For notational convenience, we denote $\overline{K} = \min\{K,\lceil N/4\rceil\}$ and $\kappa = \min\{K,N/4\}$. 

The lower bound $R_{\sf lower}(K,N,M)$ is an intersection of half planes. The corner points of $R_{\sf lower}(K,N,M)$ are characterized by the set $\Omega = \{(\omega_k,R_{\sf lower}(K,N,\omega_k)):k\in[0:\overline{K}]\}$, where 
\begin{IEEEeqnarray*}{rCl}
\omega_k &=& \begin{cases}
N & \text{ if } k=0, \\
\frac{N\left(1-\frac{1}{N}\right)^k}{N+(k+1-N)\left(1-\frac{1}{N}\right)^k} & \text{ if } k\in[\overline{K}-1], \\
0 & \text{ if } k=\overline{K}.
\end{cases}
\end{IEEEeqnarray*}
It can be checked that for all $k\in[\overline{K}-1]$, the two lines 
\begin{IEEEeqnarray*}{rCl}
y &=& (1-(1-1/N)^k)(N-k x),  \\
y &=& (1-(1-1/N)^{k+1})(N-(k+1) x)
\end{IEEEeqnarray*}
intersect at $x=\omega_k$. 

\begin{figure}[t!]
\begin{center}
\includegraphics[scale=0.5]{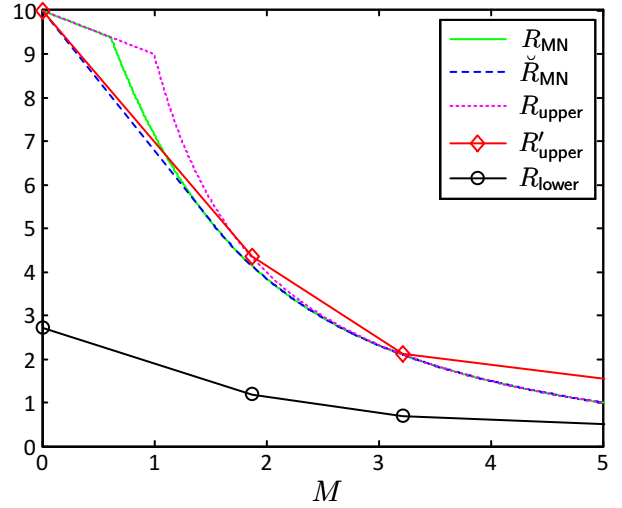}
\end{center}
\vspace{-0.2cm}
\caption{Plots of various bounds for $(K,N)=(15,10)$ and $M\in[0,5]$.}
\label{fig:N10K15}
\vspace{-0.2cm}
\end{figure}

\begin{figure*}[!t]
\normalsize
\begin{IEEEeqnarray*}{rCl}
&=& \frac{\left[N+(k-N)\left(1-\frac{1}{N}\right)^k\right]\left[N+(k-N+1)\left(1-\frac{1}{N}\right)^k\right]}{\left[N\left(1-\frac{1}{N}\right)^k\right]\left[1-\left(1-\frac{1}{N}\right)^k\right]\left[N+(-N+1)\left(1-\frac{1}{N}\right)^k\right]} \\
&=& \frac{\left[1-\left(1-\frac{1}{N}\right)^k+\frac{k}{N}\left(1-\frac{1}{N}\right)^k\right]\left[1-\left(1-\frac{1}{N}\right)^{k+1}+\frac{k}{N}\left(1-\frac{1}{N}\right)^k\right]}{\left(1-\frac{1}{N}\right)^k \left[1-\left(1-\frac{1}{N}\right)^k\right]\left[1-\left(1-\frac{1}{N}\right)^{k+1}\right]} \\
&=& \frac{1}{\left(1-\frac{1}{N}\right)^k}\left[1 + \frac{\frac{k}{N}\left(1-\frac{1}{N}\right)^k}{1-\left(1-\frac{1}{N}\right)^{k}}+\frac{\frac{k}{N}\left(1-\frac{1}{N}\right)^k}{1-\left(1-\frac{1}{N}\right)^{k+1}}+ \frac{\left(\frac{k}{N}\left(1-\frac{1}{N}\right)^k\right)^2}{\left[1-\left(1-\frac{1}{N}\right)^{k}\right]\left[1-\left(1-\frac{1}{N}\right)^{k+1}\right]}\right] 
\end{IEEEeqnarray*}
\hrulefill
\vspace*{0pt}
\end{figure*}

Next, we relax the upper bound $\breve{R}_{\sf upper}(K,N,M)$ by the following piecewise-linear bound resulting from $\{\omega_k:k\in[0:\overline{K}]\}$: 
\begin{IEEEeqnarray*}{ll}
& R'_{\sf upper}(K,N,M) \\
&:= (1-\theta)R_{\sf upper}(K,N,\omega_{k}) + \theta R_{\sf upper}(K,N,\omega_{k-1}),
\end{IEEEeqnarray*}
where $M = (1-\theta)\omega_{k} + \theta\omega_{k-1}$ for some $\theta\in [0,1)$, $k\in[\overline{K}]$. Note that $R'_{\sf upper}(K,N,M)
= R_{\sf upper}(K,N,M)$ for all $M\in\{\omega_k:k\in[0:\overline{K}]\}$. In Figure \ref{fig:N10K15} we provide an example with $(K,N)=(15,10)$ summarizing the various bounds used in the analysis.

Then, for each segment $[\omega_{k},\omega_{k-1})$, $k\in[\overline{K}]$, the ratio $\frac{R'_{\sf upper}(K,N,M)}{R_{\sf lower}(K,N,M)}$ is a linear-fractional function with respect to $M$, and thus it is quasiconvex \cite{Boyd:04}. A quasiconvex function has the property that the value of the function on a segment does not exceed the maximum of its values at the endpoints. That is to say, it suffices to check whether $\frac{R'_{\sf upper}(K,N,M)}{R_{\sf lower}(K,N,M)}< 4.7$ for all $M\in\{\omega_0,\omega_1,\cdots,\omega_{\overline{K}}\}$. 

First, it is clear that we have 
\begin{IEEEeqnarray*}{rCl}
\lim_{M\uparrow N}\frac{R'_{\sf upper}(K,N,M)}{R_{\sf lower}(K,N,M)} &=& 1.
\end{IEEEeqnarray*}
Next, we have 
\begin{IEEEeqnarray*}{rCl}
\frac{R'_{\sf upper}(K,N,0)}{R_{\sf lower}(K,N,0)} &=& \frac{\min\{K,N\}}{(1-(1-1/N)^{\overline{K}})N} \\
&\le& \frac{4\kappa/N}{1-(1-1/N)^{\kappa}} \\ 
&\overset{(a)}{\le}& 4\cdot \frac{\kappa/N}{1-e^{-\kappa/N}} \\
&\overset{(b)}{\le}& \frac{1}{1-e^{-1/4}} \approx 4.521, 
\end{IEEEeqnarray*}
where $(a)$ follows since $(1-1/z)^z \le e^{-1}$ for all $z > 1$ and $(b)$ follows since $\psi(z)=\frac{z}{1-e^{-z}}$ is an increasing function and $\kappa/N\le 1/4$. 

As for $k\in[\overline{K}-1]$, we have 
\begin{IEEEeqnarray*}{ll}
& \frac{R'_{\sf upper}(K,N,\omega_k)}{R_{\sf lower}(K,N,\omega_k)} \\
&\le\frac{\frac{N-\omega_k}{\omega_k}}{(1-(1-1/N)^k)(N-k\omega_k)} \\
&= \text{(see the top of the current page)} \\
&\le \frac{1}{\left(1-\frac{1}{N}\right)^k}\left(1 + \frac{\frac{k}{N}\left(1-\frac{1}{N}\right)^k}{1-\left(1-\frac{1}{N}\right)^{k}}\right)^2 \\
&\overset{(a)}{=} e^{z}\left(1 + \frac{1}{\ln(1-1/N)^{-N}}\frac{z}{e^{z}-1}\right)^2 \\
&\overset{(b)}{\le} e^{z}\left(1 + \frac{z}{e^{z}-1}\right)^2  \\
&\overset{(c)}{\le} \left. e^z\left(1 + \frac{z}{e^z-1}\right)^2 \right|_{z=-\frac{N}{4}\ln(1-1/N)}, 
\end{IEEEeqnarray*}
where $(a)$ follows by a change of variable $z=-k\ln(1-1/N)$, $(b)$ follows since $z\ge 0$ and $(1-1/N)^{-N}\ge e$ for all $N > 1$, $(c)$ follows since $\phi(z) = e^{z}\left(1 + \frac{z}{e^{z}-1}\right)^2$ is an increasing function$^{\S}$\footnote{$^{\S}$ It can be verified that the function $\phi(z) = e^{z}\left(1 + \frac{z}{e^{z}-1}\right)^2$, $z\ge0$, is an increasing function by showing that its first derivative is nonnegative.} and $z\le -\frac{N}{4}\ln(1-1/N)$ (since $k \le \overline{K}-1 \le N/4$). Finally, since $-N\ln(1-1/N)$ is a decreasing function of $N$ and $N\ge 5$, we have 
\begin{IEEEeqnarray*}{rCl}
\frac{R'_{\sf upper}(K,N,\omega_k)}{R_{\sf lower}(K,N,\omega_k)} &\le& \left. e^z\left(1 + \frac{z}{e^z-1}\right)^2 \right|_{z=-\frac{5}{4}\ln(1-1/5)} \\
&=& \left. \nu^\nu\left(1 + \frac{\nu\ln\nu}{\nu^\nu-1}\right)^2 \right|_{\nu=\frac{5}{4}} \\
& \approx & 4.607.
\end{IEEEeqnarray*}

%
%

\section*{Appendix: Proof of Theorem \ref{thm:subset_ind}}
For the single-user caching problem, a single-letter characterization of the optimal memory--rate tradeoff for the average case can be found in \cite[Chapter 3.7.3]{CYWang:15}. For the considered setup in Theorem \ref{thm:subset_ind}, the optimal memory--rate tradeoff can be expressed as 
\begin{IEEEeqnarray*}{rCl}
R^\star_{\sf avg}(M) &=& \min H(B_Y|V,Y),
\end{IEEEeqnarray*}
where the minimum is over all conditional pmfs $p_{V|B}$ such that $V\markov B \markov Y$ form a Markov chain and $I(B;V) \le M$.

{\em (Converse.)} Let $M\ge 0$ be fixed. Consider any conditional pmf $p_{V|B}$ such that $V\markov B \markov Y$ form a Markov chain and $I(B;V) \le M$. Recall that $B=(B_1,\cdots,B_N)$. Then, for all $n\in[N]$, we have 
\begin{IEEEeqnarray*}{rCl} 
M &\ge& I(B;V)  \\
&\ge& I(B_{[n]};V|B_{[n+1:N]}) \\
&=& H(B_{[n]}) - H(B_{[n]}|V,B_{[n+1:N]}) \\
\label{eq:cond_subsetind}
&=& n - H(B_n|V,B_{[n+1:N]}) - H(B_{[n-1]}|V,B_{[n:N]}) \IEEEyesnumber. 
\end{IEEEeqnarray*}
Now we show that 
\begin{IEEEeqnarray*}{rCl}
H(B_Y|V,Y) &\ge& \sum_{n=1}^N (s_n-s_{n+1}) \left(n-M\right)^+. 
\end{IEEEeqnarray*}
First, we have 
\begin{IEEEeqnarray*}{ll}
& H(B_Y|V,Y) \\
&= \sum_{y\in\mathcal{P}([N])} p_Y(y) H(B_y|V) \\
&\overset{(a)}{\ge} \sum_{n=1}^N s_n H(B_n|V,B_{[n+1:N]}), 
\end{IEEEeqnarray*}
where $(a)$ follows by recursively applying the inequality 
\begin{IEEEeqnarray*} {ll}
& \sum_{y\in\mathcal{P}([N])} p_Y(y) H(B_y|V,B_{[n+1:N]}) \\
&\ge s_nH(B_n|V,B_{[n+1:N]}) \\
&\quad + \sum_{y\in\mathcal{P}([N])} p_Y(y) H(B_{y}|V,B_n,B_{[n+1:N]}), 
\end{IEEEeqnarray*}
in the order $N,N-1,\cdots,1$. Next, $H(B_Y|V,Y)$ can be further lower bounded as  
\begin{IEEEeqnarray*} {ll}
& H(B_Y|V,Y) \\
&\ge \sum_{n=1}^N s_n H(B_{n}|V,B_{[n+1:N]}) \\
&= s_N H(B_{N}|V) + \sum_{n=1}^{N-1} s_n H(B_{n}|V,B_{[n+1:N]}) \\
&\overset{(a)}{\ge} s_N \left(N-M-H(B_{[N-1]}|V,B_{N})\right)^+ \\
& \hspace{0.5cm} + \sum_{n=1}^{N-1} s_n H(B_{n}|V,B_{[n+1:N]}) \\
&\overset{(b)}{\ge} s_N \left(N-M\right)^+ - s_NH(B_{[N-1]}|V,B_{N}) \\
& \hspace{0.5cm} + \sum_{n=1}^{N-1} s_n H(B_{n}|V,B_{[n+1:N]}) \\
&= s_N \left(N-M\right)^+ + \sum_{n=1}^{N-1} (s_n-s_N) H(B_{n}|V,B_{[n+1:N]}) \\
&= s_N \left(N-M\right)^+ + (s_{N-1}-s_N) H(B_{N-1}|V,B_{N}) \\
& \hspace{0.5cm} + \sum_{n=1}^{N-2} (s_n-s_N) H(B_{n}|V,B_{[n+1:N]}) \\
&\overset{(c)}{\ge} s_N \left(N-M\right)^+ \\
& \hspace{0.5cm} + (s_{N-1}-s_N) \left(N-1-M - H(B_{[N-2]}|V,B_{[N-1:N]})\right)^+ \\
& \hspace{0.5cm} + \sum_{n=1}^{N-2} (s_n-s_N) H(B_{n}|V,B_{[n+1:N]}) \\
&\overset{(d)}{\ge} s_N \left(N-M\right)^+ + (s_{N-1}-s_N) \left(N-1-M \right)^+ \\
& \hspace{0.5cm} - (s_{N-1}-s_N) H(B_{[N-2]}|V,B_{[N-1:N]}) \\
& \hspace{0.5cm} + \sum_{n=1}^{N-2} (s_n-s_N) H(B_{n}|V,B_{[n+1:N]}) \\
&= s_N \left(N-M\right)^+ + (s_{N-1}-s_N) \left(N-1-M \right)^+ \\
& \hspace{0.5cm} + \sum_{n=1}^{N-2} (s_n-s_{N-1}) H(B_{n}|V,B_{[n+1:N]}), \\
\end{IEEEeqnarray*}
where $(a)$ and $(c)$ follow from \eqref{eq:cond_subsetind} and $H(B_{n}|V,B_{[n+1:N]})\ge0$ with $n=N$ and $n=N-1$, respectively, and $(b)$ and $(d)$ follow since $(u-v)^+\ge(u)^+-v$ for all $v\ge 0$.
At this point, it is clear that we can apply the same argument for another $N-2$ times and arrive at 
\begin{IEEEeqnarray*}{rCl}
\label{eq:lower_subsetind}
H(B_Y|V,Y) &\ge& \sum_{n=1}^N (s_n-s_{n+1}) \left(n-M\right)^+, \IEEEyesnumber
\end{IEEEeqnarray*} 
where $s_{N+1}=0$.

{\em (Achievability.)} Note that the lower bound \eqref{eq:lower_subsetind} is equivalent to saying that 
\begin{enumerate}
\item if $M \ge N$, then $H(B_Y|V,Y) \ge 0$, and 
\item if $n-1 \le M < n$ for some $n\in[N]$, then  
\begin{IEEEeqnarray*}{ll}
& H(B_Y|V,Y) \\
&\ge s_n\left(n-M\right) + \sum_{j=n+1}^N s_{j} \\
&= \sum_{\substack{ y\in\mathcal{P}[N]  \\  \text{ s.t. } n\in y}} p_Y(y) \left(n-M\right) + \sum_{y\in\mathcal{P}[N]} p_Y(y)H(B_{y}|B_{[n]}).
\end{IEEEeqnarray*}
\end{enumerate}
Therefore, for all $M\in\{0\}\cup[N]$, setting $V=(B_{1},\cdots,B_{M})$ makes \eqref{eq:lower_subsetind} hold with equality. Since the rest of the memory--rate tradeoff can be achieved by time sharing, the achievability is established.

\section*{Acknowledgement}
This work has been supported in part by the European ERC
Starting Grant 259530-ComCom.

\bibliographystyle{IEEEtran}
\bibliography{IEEEabrv,References}

\end{document}